\documentclass[twocolumn,preprintnumbers,amsmath,amssymb,nofootinbib,APS]{revtex4}

\usepackage{dcolumn}
\usepackage{bm}

\usepackage{amsmath,amsfonts,amssymb}
\usepackage{graphicx}
\usepackage{hyperref}

\newcommand{\be}{\begin{equation}}
\newcommand{\ee}{\end{equation}}
\newcommand{\matrice}{\begin{pmatrix}}
\newcommand{\ematrice}{\end{pmatrix}}

\newcommand{\bea}{\begin{eqnarray}}
\newcommand{\eea}{\end{eqnarray}}

     \let\d=\delta
     \let\th=\theta  \let\k=\kappa \let\l=\lambda
\let\m=\mu    \let\n=\nu            
     \let\ph=\phi 
    
 \let\D=\Delta    
         
  \let\eps=\epsilon

\let\==\equiv
\let\io=\infty

\let\p=\partial

\def\PPP{{\cal P}} 
\def\CC{{\cal C}}

\def\to{\rightarrow}
\def\la{\left\langle}
\def\ra{\right\rangle}

\def\Tr{{\rm Tr}\,}
\def\la{\langle}
\def\ra{\rangle}

\begin{document}

\title{{\bf Fractal properties of quantum spacetime}}
\author{Dario Benedetti}\email{dbenedetti@perimeterinstitute.ca}
\affiliation{ {\footnotesize Perimeter Institute for Theoretical Physics,
31 Caroline St. N, N2L 2Y5, Waterloo ON, Canada}}


\begin{abstract}
We show that in general a spacetime having a quantum group symmetry has also a scale dependent fractal dimension which deviates from its classical value
at short scales, a phenomenon that resembles what observed in some approaches to quantum gravity.
In particular we analyze the cases of a quantum sphere and of $\k$-Minkowski, the latter being relevant
in the context of quantum gravity.

\end{abstract}


\maketitle

In the quest for new physics at the Planck scale the idea that spacetime might become noncommutative \cite{connes,doplicher,szabo}
has gained a lot of attention, in particular for its potential phenomenological implications \cite{amelino-waves}.
Whether such idea is supposed to be taken as a starting point for the construction a quantum theory of gravity (see for example \cite{doplicher}) or can be derived from it (for 
example \cite{freidel}) there are several reasons why it could play a role at the Planck scale.
Somehow postponing the issue of a more complete and fundamental theory most of the efforts in the literature have gone on the study
of noncommutative versions of flat spacetime, which naively might be thought as a ground state of the full theory
of quantum gravity.

On the other hand constructive approaches to quantum gravity, such as
causal dynamical triangulations (CDT) \cite{ajl-rec} and exact renormalization group (ERG) \cite{reuter-rev}, 
which make no use of postulated new physics, have something interesting to say about Planck scale properties of spacetime.
It is somehow surprising to see that apparently very different approaches give rise to very similar
results as it is the case for the spectral dimension of spacetime:
both  in CDT  \cite{renate-ds} and in ERG \cite{reuter-ds}  evidence has been given for the emergence of a (ground state) spacetime with fractal properties
such as the effective (spectral) dimension $d_s$ varying from a classical value $d_s=4$ at large scales down
to $d_s=2$ at short scales.
It is a legitimate and interesting question to ask whether such a fractal nature of spacetime is compatible with the
expectation of some sort of noncommutativity.

An appealing realization of noncommutativity is that in which spacetime remains
maximally symmetric but the Lie group of symmetries is deformed into a quantum group (as in \cite{MajidRuegg}),
a deformation also favoured by general arguments on the possible non-locality of a final quantum theory of gravity \cite{arzano},
and which constitutes a solid realization of the so-called Doubly Special Relativity \cite{amelino-dsr,kowalski}.
Research in this area in still at an early stage and a complete formulation of quantum field theory based on a quantum group symmetry is still lacking,
but some proposals have been put forward for the construction of the corresponding Fock space (see for example \cite{ab-08} and references therein).
Here we explore the geometrical properties of such type of spacetimes by calculating the spectral dimension associated with them.
In order to do so we adopt a group theoretical construction that suits well to the quantum group formalism.
We find for the noncommutative spacetimes considered a result qualitatively similar to that found in CDT and ERG, {\it i.e.}
a scale dependent spectral dimension which reaches its classical value only at large scales.

{\it Spectral dimension} --
A possible way to study the geometry of a Riemannian manifold $M$ with metric $g_{\m\n}$ is via the spectral theory of the scalar Laplacian $\D=-g^{\m\n}\nabla_\m\nabla_\n$,
where $\nabla_\m$ is the covariant derivative.
To such an operator it can be associated a heat kernel, $i.e.$ a function $K(x,y;s)$ on $M \times M\times \mathbb{R}_+$ which solves the heat equation
\be \label{heat-eq}
\p_s K(x,y;s) + \D_x K(x,y;s) =0\ ,
\ee
with the initial condition $K(x,y;0_+) = \d(x-y)/\sqrt{g(x)}$.

It is a well known result that the (normalized) trace of the heat kernel has the following expansion \cite{vassilevich}
\be \label{HK-expansion}
\Tr K = \frac{\int d^n x \sqrt{g(x)} K(x,x;s)}{\int d^n x \sqrt{g(x)}} \sim \tfrac{1}{(4\pi s)^{n/2}} \sum_{i=0}^{+\io} a_i s^i\ ,
\ee
where the coefficients are metric-dependent invariants which can be calculated via recursion formulas, with $a_0=1$.
It is then possible to define the notion of \emph{spectral dimension} by the formula
\be \label{d_s}
d_s \= -2 \frac{\p \log \Tr K}{\p \log s}
\ .
\ee
On flat space $a_i\= 0$ for $i\geq 1$ and so we recover $d_s= n$.
On a general classical curved space $d_s=n$ only at small $s$, while deviations occur at large $s$ due to the curvature.
Since we can identify the diffusion time $s$ with the scale at which we probe the manifold, when applying the classical expansion (\ref{HK-expansion}) to our spacetime we 
should take $s$ to be small compared to the characteristic
dimension of the space, but still large compared to the Planck scale, else this formula will
not be valid anymore because of the metric fluctuations, as suggested by the results in \cite{renate-ds,reuter-ds}.

The usefulness of such a definition is in providing an operational notion of dimension, which is a valuable
alternative to the maybe more famous Hausdorff dimension associated to the scaling exponent of the volume of a ball.

The solution of (\ref{heat-eq}) is given by $K=\la x| e^{-s \D}|y \ra$, or in terms of eigenvalues $\l_j$ and eigenfunctions $\ph_j(x)$ of $\D$
\be \label{eigenf-heat}
K(x,y;s)=\sum_j e^{-\l_j s} \ph_j(x) \ph^*_j(y)\ ,
\ee
where it has to be understood that the spectrum might be continuum and in such case the sum would be replaced by an integral.
In flat spacetime for example we have
\be \label{flat}
K_{\text{flat}}(x,y;s) = \int \tfrac{d^n p}{(2\pi)^n} e^{-p^2 s} e^{i p\cdot (x-y)} = \tfrac{e^{-\frac{|x-y|^2}{4 s}}}{(4\pi s)^{n/2}}\ .
\ee

We now want to generalize this notion to a noncommutative space of the kind associated to a quantum group symmetry.
In such a space it is natural to define the Laplacian from the quadratic Casimir of the quantum group,
in analogy to the general construction on homogeneous spaces \cite{camporesi}.
For example, in the case of a flat Euclidean space $E^n\sim ISO(n)/SO(n)$ we find that the spectrum of the Laplacian is given by the first Casimir $\CC_1=P_\m P^\m$ in the 
irreducible scalar representations,
thus recovering (\ref{flat}).
We can follow this route in a straightforward way for the case of a quantum group, and in particular for a quantum deformation of the Poincar\'e group,
as we will now show.

{\it A toy example: the sphere vs the quantum sphere} --
To illustrate the idea, it is useful to look at a simple example first.
Following \cite{camporesi} we write the heat kernel on a homogeneous space as an integral of the heat kernel on the symmetry group $G$ over the isotropy
group $H$
\be \label{quotient-heat}
K_{G/H}(x,y;s)=\int_H K_G(e,gh;s) dh\ ,
\ee
where $y=g x$, $g\in G$ and $K_G$ can be obtained by a character expansion
\be \label{group-heat}
K_G(g;s)\equiv K_G(e,g;s) = \frac{1}{V_G}\sum_j d_j \chi_j(g) e^{-s \CC(j)}\ ,
\ee
where the sum is over all the irreducible representations of $G$, $d_j$ is their dimension, $\chi_j(g)$ is the character of $g\in G$ in the representation $j$, $\CC(j)$ the value of 
the Casimir
in that representation, and $V_G$ the volume of $G$. 
Plugging (\ref{group-heat}) into (\ref{quotient-heat}) one finds that the integration restricts the summation to be only over the spherical representations of $G$ with respect to $H$,
$i.e.$ those which contain the singlet of $H$.

\begin{figure}[h]
\begin{center}
\includegraphics[width=8cm]{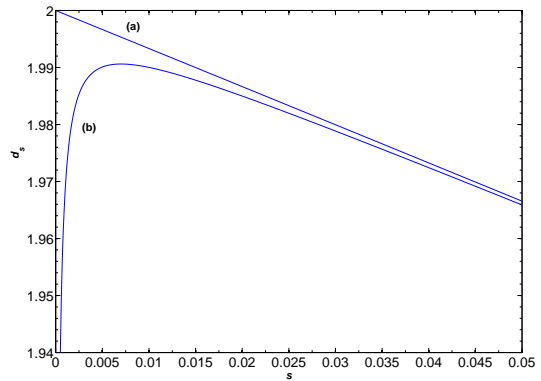} 
\end{center}
\caption{\footnotesize \mbox{\bf (a):} The spectral dimension of a unit sphere  $SU(2)/U(1)$.
\mbox{\bf (b):} The case of a quantum sphere $SU_q(2)/U(1)$, for $z=0.01$.}
\label{sfere}
\end{figure}

Our first example, before moving to quantum spaces, is the classical (unit) two-sphere $S^2$ considered as $SU(2)/U(1)$.
Using  (\ref{group-heat}) and (\ref{quotient-heat}) one finds the expression
\be
K_{S^2}(\th;s)=\frac{1}{4\pi}\sum_{l=0}^{\io}(2l+1) P_l(\cos\th) e^{-s l (l+1)}\ ,
\ee
where $P_l(x)$ are the Legendre polynomials and $\CC(l)=l(l+1)$, and which is of course equivalent to the expression (\ref{eigenf-heat}).
Taking the trace is trivial, and we find
\be \label{sfera-heat}
\Tr K_{S^2}=\sum_{l=0}^{\io}(2l+1) e^{-s l (l+1)}\ .
\ee
Finally we can use formula (\ref{d_s}) to get the spectral dimension, which we plot in Fig.~\ref{sfere}a. Note that the value $d_s=2$ is reached
exactly at $s=0$ and away from that it decreases due to the curvature.

Consider now replacing the group $SU(2)$ by the quantum group $SU_q(2)$ for real $q$, which is generated by the operators $J_+$, $J_-$ and $J_3$ obeying the commutation
relations
\be
[J_3,J_{\pm}]=\pm J_{\pm}\ , \hspace{.7cm} [J_+,J_-]=\frac{\sinh (z J_3)}{\sinh (z/2)}\ ,
\ee
where $z=\ln q$. Such generators belong to the quantum Hopf algebra $U_q(su(2))$ whose representations are well known (see for example \cite{biedenharn})
and parallel (for real $q$) those of $su(2)$, in the sense that for every $j=0,\frac{1}{2},1,...$ the Hopf algebra  $U_q(su(2))$ has a $2j+1$-dimensional representation
$\{|j,m\ra,m=-j,-j+1,...,j\}$ with $J_3|j,m\ra=m|j,m\ra$ (the action of $J_\pm$ and the coalgebra structure are different from those of $su(2)$ but we don't need them here).
The Casimir in the representation $j$ is given by
\be \label{qsfera-casimir}
\CC(j)=\frac{\cosh(z(2j+1)/2)-\cosh(z/2)}{2 \sinh^2(z/2)}\ .
\ee

The above steps for the case of $SU(2)$ can be repeated for $SU_q(2)$, in particular the integration over $U(1)$ restricts the sum over representations to
only those with integer $j$, $i.e.$ those containing the singlet of $U(1)$ which correspond to $m=0$.
We only need to replace in (\ref{sfera-heat}) the standard $su(2)$ Casimir with (\ref{qsfera-casimir}).

Again the spectral dimension can be computed (numerically) using (\ref{d_s}) and the result is plotted in Fig.~\ref{sfere}b.
Clearly the behaviour is the standard one for large $s$ but it deviates sensibly as $s$ decreases, with $d_s$ never reaching the value $d=2$
and going instead down to zero.
We can think of this phenomenon as a signature of the fuzziness of the quantum sphere, or of fractal behaviour at short scales.

{\it $\k$-Minkowski} --
$\k$-Poincar\'e was derived in \cite{luk-92} as a particular contraction of the quantum anti-de Stitter algebra $U_q(O(3,2))$ in which the anti-de Sitter radius $R$ goes to infinity 
while 
$R \ln q=\k^{-1}$ is a real number which is held fixed and finite. As explained in \cite{amelino-artem} such limit might be of relevance for a
theory of quantum gravity.
The result of such contraction is another Hopf algebra
which in terms of the generators of translations $P_\m$, rotations $M_j$ and boosts $N_j$  (as usual Greek indices run from 0 to 3
while Latin indices run from 1 to 3, and repeated indices are summed over) has the deformed algebra relations
\bea \label{old-basis}
& [N_i,P_j]=\d_{ij} \k \sinh \frac{P_0}{\k}\ , \\ \nonumber
& [N_i,N_j]=-\eps_{ijk}(M_k \cosh \frac{P_0}{\k}-\frac{1}{4\k^2}P_k P_l M_l)\ ,
\eea
the other commutators being as in undeformed Poincar\'e.
As shown in \cite{k-reps} hermitian irreducible representations of the Poincar\'e algebra with $\CC_1=P_\m P^\m\geq 0$ can be lifted to hermitian irreducible representations of $\k$-Poincar\'e
with $\CC^\k_1=(2\k \sinh \frac{P_0}{2\k})^2-\vec P^2\geq 0$,
and the latter reduce in the $\k\to\io$ limit to the undeformed ones.

$\k$-Minkowski spacetime was introduced in \cite{MajidRuegg} as the space which is dual to the translation sector of $\k$-Poincar\'e algebra and on which the whole $\k$-
Poincar\'e algebra
acts covariantly, and as such is a subgroup of the so-called $\k$-Poincar\'e group \cite{zakr}. It turns out to be a noncommutative spacetime with coordinates $\hat x^{\mu}$
satisfying the relations
\be
[\hat x^0,\hat x^j]=\frac{i}{\k}\hat x^j\ , \hspace{1cm} [\hat x^i,\hat x^j]=0\ .
\ee
 
Following \cite{MajidRuegg} it is convenient to introduce a new basis for $\k$-Poincar\'e by defining new boost generators
\be
N^b_j=N_j e^{-\frac{P_0}{2\k}} -\frac{\eps_{jkl}}{2\k} M_k P_l e^{-\frac{P_0}{2\k}}\ ,
\ee
such that the bicrossproduct structure
of $\k$-Poincar\'e $\PPP_\k=U(so(3,1) \triangleright\!\!\!\!\blacktriangleleft T$ becomes evident, with generators of rotations and boost
forming the standard Lorentz algebra
and with deformed action of $U(so(3,1)$ on $T$ given by the remaining commutators
\bea
&[N^b_i,P_0]=e^{-\frac{P_0}{2\k}} P_i\ ,\\ \nonumber
& [N^b_i,P_j]=\d_{ij} e^{-\frac{P_0}{2\k}} (\k \sinh \frac{P_0}{\k} +\frac{1}{2\k} \vec P^2)-\frac{1}{2\k}e^{-\frac{P_0}{2\k}}P_i P_j\ .
\eea
One nice consequence of the bicrossproduct structure is that the dual $P^*_\k$ possesses the same structure, $i.e.$  $P^*_\k = T^* \blacktriangleright\!\!\!\triangleleft \mathbb{C} 
(SO(3,1))$,
and so we can think of  $\k$-Minkowski as the homogeneous space $P^*_\k/SO(3,1)$, and this justifies us in applying the previous formalism
for evaluating the trace of the heat kernel on $\k$-Minkowski.

Before doing that we have to switch to Euclidean signature in order to make sense of our definition of effective dimension, but this constitutes no problem, it just amounts to
the substitution (see for example \cite{Luk-eucl}) $P_0\to i P_0$, $\k\to i \k$.
When applied to the first Casimir of the algebra such substitution yields
\be \label{k-casimir}
\CC^\k_1=(2\k \sinh \frac{P_0}{2\k})^2+\vec P^2\ ,
\ee
in agreement with the most intuitive extension of the two- and three-dimensional cases of \cite{firenze}.

Next we also have to note that any function of $\CC^\k_1$ is still a valid Casimir\footnote{Having an extra parameter which is dimensionful  we can construct arbitrary
functions with mass-squared dimension, the only restriction being given by the limit $\k\to\io$, for which we ought to recover the standard Casimir.}.
To select one unique expression we can make appeal to the existing theory of differential calculus on $\k$-Minkowski \cite{diff-calc} (see also \cite{freidel-noether} for recent applications to 
quantum field theory
on $\k$-Minkowski) and compare our group theoretical construction with the Laplacian defined via such differential calculus.
We find that in the basis we have chosen the eigenvalues of the Laplacian are given by
\be \label{k-eig}
M^2(p)=\CC^\k_1(p) (1+\frac{\CC^\k_1(p)}{4 \k^2})\ .
\ee

\begin{figure}[ht]
\centering
\vspace*{13pt}
\includegraphics[width=7cm]{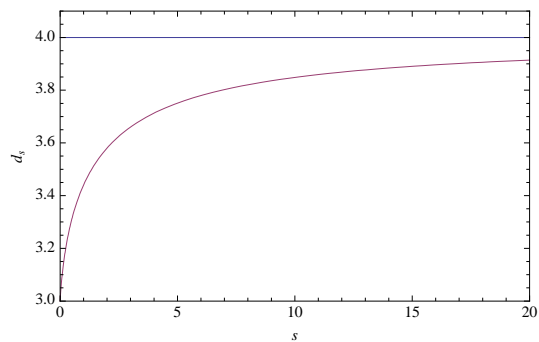}
\vspace*{13pt}
\caption{\footnotesize A plot of the spectral dimension $d_s$ of $\k$-Minkowski space for $\k=1$ as function of the diffusion time $s$.
For comparison we plot also the constant behaviour of the spectral dimension of classical Minkowski space ($d_s=4$).}
\label{spectralDim}
\end{figure}

We can now use $M^2(p)$ as the Casimir eigenvalue, and write down the following formula for $\k$-Minkowski
\be \label{k-kernel}
\Tr K_q = \int \frac{d\m( p)}{(2\pi)^4}\ e^{-s M^2 (p)}\ ,
\ee
where we have also used the $\k$-deformed Lorentz invariant measure $d\m(p)\equiv e^{\tfrac{3p_0}{2\k}} d^4p$.
Finally from (\ref{d_s}) we obtain the spectral dimension of (the Wick-rotated) $\k$-Minkowski space.
The integration cannot be done analytically, but numerically it poses no problems and we can plot the result as for example in Fig.~\ref{spectralDim}.
The limiting values at $s\to\io$ and $s\to 0$ can be obtained analytically by respectively taking the limits of small and large $p_0/\k$ for the integrand, obtaining
\be \label{k-d_s}
d_s=\begin{cases}  4 & \text{for $s\to \io$}\\ 3 & \text{for $s\to 0$.} \end{cases}
\ee

The behaviour in (\ref{k-d_s}), our main result, is qualitatively similar to those in \cite{renate-ds,reuter-ds} with the main difference being the short scale behaviour
leading to an effective dimension $d_s=3$ in our case rather than $d_s=2$ as in  \cite{renate-ds,reuter-ds}.
The fact that the effective dimensionality departs from an integer value, and from the topological dimension in particular, is a typical signature
of fractal geometry. The meaning of fractal in the context of noncommutative geometry is actually a largely unexplored subject, 
but certainly the behaviour found for the spectral dimension
can be interpreted qualitatively as a defining property of fractal nature.

It is important to note also that the result obtained is independent of the choice of momentum basis, as that 
 would just amount
to a change of variables in (\ref{k-kernel}). It is less trivial to check the independence on the choice of basis in the Lorentz sector, as this affects the integration measure and the 
choice
of Casimir function $M^2(p)$. Here we just note that using the original basis (\ref{old-basis}), which requires a trivial measure, and using $M^2(p)=\CC^\k_1(p)$ we find the same
result as in (\ref{k-d_s}).

The result (\ref{k-d_s}) can also be understood by noticing that the dispersion relation (\ref{k-casimir}) looks like that associated to a finite difference operator
(along the time direction).
Such an interpretation of the $\k$-deformed Klein-Gordon operator was known since
the early days of $\k$-Poincar\'e \cite{lattice}.
In light of this analogy one might then think of the diffusion process being trapped  at short diffusion-times $s$ in a countinuous three-dimensional slice and that only at large 
scales
the discreteness in time $t$ would become irrelevant and thus look like an additional continuous dimension.
On the other hand the analogy is purely formal as one should notice that the finite difference operator acts along the imaginary axis and that there is actually no discretization
of time in the $\k$-Minkowski construction ($t$ can take any value).
For these reasons we prefer to think of  (\ref{k-d_s}) as a result of the noncommutativity of spacetime at short scale, with the consequent uncertainty relations
that would allow one to precisely determine three space coordinates but not the fourth (time).

{\it Conclusions} -- 
We have shown how the result of \cite{renate-ds,reuter-ds} about the dynamical dimensional reduction at short scales can be reproduced, in its qualitative aspect, by a noncommutative spacetime
with quantum group symmetry.
In light of the comment above it is tempting to conjecture that the value of the spectral dimension in the far ultraviolet limit is generally given by the dimension
of the maximal commutative subspace. If that turns out to be true, at least within certain hypothesis (for the sphere above it's not true!),
then it would be easy to construct a spacetime whose spectral dimension goes to 2 in the UV, thus paralleling the result in \cite{renate-ds,reuter-ds} also quantitatively.
Less trivial is to identify the associated quantum group and thus prove such a conjecture. We hope to come back to this issue in the near future.

{\it Acknowledgments} -- 
I would like to thank  Michele Arzano and Simone Speziale for comments.   
Research at Perimeter Institute for Theoretical Physics is supported in part by the Government of Canada through NSERC and by the Province of Ontario through MRI.

\bibliographystyle{unsrt}

\end{document}